%% file: ProcCONF16.tex
\documentclass[epj]{webofc}
\usepackage[varg]{txfonts}   
\usepackage{dsfont}
\usepackage{upgreek}
\usepackage{microtype}

\usepackage[
    toc,
    acronym,
    nomain,
    style=long,
    nonumberlist,
    nowarn,
    ]{glossaries}
\makeglossaries
\glsdisablehyper
\input{glossaries}

\wocname{EPJ Web of Conferences}
\woctitle{CONF12}
\begin{document}
\selectlanguage{english}
\title{Charming quasi-exotic open-flavor mesons}
%
%

\author{Thomas Hilger\inst{1,2}\fnsep\thanks{\email{thomas.hilger@uni-graz.at}}
        \and
        Andreas Krassnigg\inst{1}
}

\institute{Institute of Physics, University of Graz, NAWI Graz, A-8010 Graz, Austria
\and
           Institute of High Energy Physics, Austrian Academy of Sciences, A-1050 Vienna, Austria 
}

\abstract{%
  We discuss charmed mesons in the covariant Dyson-Schwinger-Bethe-Salpeter-equation approach. 
  In particular we computed masses, leptonic decay constants, and an orbital-angular-momentum decomposition for a basic set of states. 
  We also report an efficient way to treat the two coupled quark propagator dressing functions via a single function.
}
\maketitle
\section{Introduction}
\label{intro}

Investigating the non-perturbative features of \gls{QCD}, in particular \gls{DCSB} and confinement, and their relation to the formation and appearance of the visible matter in the universe, are as central aspects of current hadron physics as the properties of matter at high temperatures and densities.
Similar appeal is attributed to the observation and an in-depth understanding of patterns of mesons with exotic quantum numbers \cite{Meyer:2010ku,Briceno:2015rlt,Geesaman:2015fha}.
Among other intersting topics, the role of \gls{DCSB} in excited and exotic meson states is of high interest and significant difference from ground states.
Hence, exotic mesons are an important object of study in order to understand the effects of \gls{DCSB} on meson properties.

Furthermore, it has been demonstrated \cite{Hilger:2009kn,Hilger:2010zb,Kampfer:2010vk,Rapp:2011zz,Hilger:2011cq,Buchheim:2014ija,Buchheim:2014eya,Buchheim:2014uda,Buchheim:2014rpa,Buchheim:2015xka,Buchheim:2015yyc,Hilger:2010zf} that open-flavor mesons are best suited to study the interrelation of \gls{DCSB} and medium modifications of mesons due to the direct coupling to the chiral condensate.
In particular, contrary to light-quark mesons, where the chiral condensate is suppressed by the light-quark mass \cite{Hilger:2010cn}, and heavy-quark mesons, where the chiral condensate only enters indirectly via higher-order terms of the operator product expansion \cite{Hilger:2012db}, for open-flavor mesons the chiral condensate's numerical impact is enhanced by the heavy-quark mass \cite{Reinders:1984sr,Hilger:2008jg,Zschocke:2011aa}.
Therefore, in particular quasi-exotic open-flavor mesons \cite{Hilger:2016efh} are expected to be extremely sensitive probes of \gls{DCSB} and its restoration at higher densities and/or temperatures.

While commonly exotic mesons are attributed to states with more degrees of freedom than simple quark-antiquark constellations, it is easy to see that on a Poincar\'e covariant level exotic quantum numbers can be realized already by virtue of quark-antiquark bilinears \cite{Burden:2002ps,Hilger:2015hka,Hilger:2016efh}.
In such a setting, quark-antiquark states are subject to an additional relative-time dgree of freedom as a result of the equivalence of space and time as well as the relativity of simultaneity.
This additional relative-time freedom lifts the quark-model restriction of $J^{\mathcal{P}\mathcal{C}}$ and allows for exotic quantum numbers via the corresponding $\mathcal{C}$, which otherwise can only be realized by additional gluon (hybrids) or quark (tetraquarks) degrees of freedom.

In the search for exotic mesons and the investigation of their properties, we explore the applicability of the combined \gls{DSBSE} approach \cite{Fischer:2006ub,Roberts:2007jh,Krassnigg:2009zh,Bashir:2012fs,Sanchis-Alepuz:2015tha,Eichmann:2016yit} in \gls{RL} truncation \cite{Eichmann:2008ae} to mesons consisting of at least one charm quark, thereby extending the study presented in \cite{Hilger:2016efh} to a much wider range of quark masses.
Within this approach, the non-perturbatively dressed quark propagator is obtained as solution to the quark \gls{DSE} and serves as an input for the quark-antiquark meson \gls{BSE} \cite{Krassnigg:2008gd,Blank:2010bp}.
The \gls{RL} truncation provides a reliable meson phenomenology for heavy-quark mesons \cite{Blank:2011ha,Popovici:2014pha,Hilger:2014nma,Krassnigg:2016hml}.
For light-quark mesons the situation is more difficult and a recent analysis suggests that, apart from obvious effects of resonant corrections to the truncation, the situation requires even further study to arrive at a definite conclusion \cite{Hilger:2015ora}.
However, already from the beginning, \gls{RL} truncation is protected via the identities it satisfies, in particular the \gls{AVWTI}, which ensures the correct realization of \gls{DCSB} in the \gls{DSBSE} approach by construction in every numerical study such as this.
It's range of applicability naturally comprises scenarios with \cite{Holl:2004fr} and without \gls{DCSB} \cite{Hilger:2015zva}.
First investigations of open-flavor mesons beyond \gls{RL} truncation have been performed in \cite{Gomez-Rocha:2015qga,Gomez-Rocha:2014vsa,Gomez-Rocha:2016cji}, where truncation effects have been studied both qualitatively and quantitatively.

A major long-standing unresolved issue is posed by the occurrence of complex conjugated pole pairs in the quark propagator, which spoil a direct and numerically stable solution of the homogeneous \gls{BSE} if the \gls{BSE}'s quark momentum domain covers these poles \cite{Bhagwat:2002tx}.
Direct extension beyond the pole threshold requires extremely demanding numerical procedures \cite{Dorkin:2013rsa,Dorkin:2014lxa} if the result is to be numerically stable \cite{Kubrak:2014ela,Fischer:2014cfa,Fischer:2014xha,Dorkin:2010ut,Dorkin:2010pb,Chang:2013nia}.
By pole threshold we refer to the maximal accessible bound state mass where the quark momentum domains still remain clear from non-analyticities.
Another possibility is to exploit the analytic properties of the eigenvalue curves at the on-shell bound state momentum to extract the bound state mass from off-shell eigenvalues \cite{Blank:2011qk}.
However, the latter procedure does not provide a bound state's \gls{BSA} and, therefore, does not allow for a sophisticated analysis of, e.\,g., leptonic decay constants or other observables.
Another approach still involves the use of inhomogeneous versions of the appropriate \gls{BSE}, which in general provides off-shell information and can be used to fit or extrapolate desired on-shell properties, including leptonic decay constants \cite{Bhagwat:2007rj,Holl:2003dq,Blank:2010sn}.

Our approach is to use the homogenous \gls{BSE} and maximize the pole threshold by exploiting the inherent Poincar\'e covariance of this approach.
Due to the non-perturbatively dressed quark propagator and the resulting strong momentum dependence of the quark mass function, in particular for infrared momenta, the quark momenta are fixed by the meson bound state momentum $P^2$ and the relative quark momentum $k^2$ only up to the momentum partitioning parameter $\eta$.
Having genuine solutions of the \gls{BSE} at hand, also allows to study leptonic decay constants and to further explore the notion of in-hadron condensates.
In \cite{UweHilger:2012uua} it has been demonstrated that this freedom can indeed extend the range of applicability for open-flavor mesons.
Later this algorithm has been utilized in an exploratory study of the $0^-$ open-flavor bound state masses and leptonic decay constants \cite{Rojas:2014aka}.

Within a Poincar\'e covariant approach the decomposition of equal-flavor and open-flavor bound state amplitudes in terms of Lorentz and Dirac structures is identical and, in principle, allows for the same analysis of $\mathcal{C}$-parity symmetries in both cases \cite{Hilger:2015ora}.
Furthermore, comparing \glspl{OAMD} of open-flavor and equal-flavor states does not reveal any qualitative differences in the nature of the states \cite{Hilger:2016efh}.
As, thus, input and output properties are indistinguishable w.\,r.\,t.\ the quark mass ingredients it is reasonable to assume, that these states and their properties are both related and connected by variation of the quark masses.

In general, nature and any theoretical treatment of mesons should be continuous w.\,r.\,t.\ variations of the quark masses.
Consequently, no meson states should appear or disappear from the spectrum when leaving equal flavor cases or ending up on an equal flavor case while varying quark masses.
This is indeed the case for sophisticated Poincar\'e covariant approaches and immediately raises the question about the nature of those states which are connected to exotic quark-antiquark states in the equal-flavor case.
Under the assumption that such open-flavor states are absent in the quark model both conceptually and actually in any solution set, we call them \emph{quasi-exotic} mesons \cite{Hilger:2016efh}.

The previous argument is particularly interesting when viewed apart from a Poincar\'e covariant approach. The non-restricted $\mathcal{C}$ symmetry property in an open-flavor state would raise the questions to which states in the equal-flavor case those states that match our quasi-exotic description could be connected and what their interpretation should be.


\section{Revisiting the quark Dyson-Schwinger equation}
\label{sct:DSE}

Given the dressed-quark propagator $S$ for quark momentum $p$ in the vacuum parametrized as
\begin{equation}
    S(p) \equiv i\; \gamma\cdot p\; \sigma_\mathrm{V}(p^2) + \sigma_\mathrm{S}(p^2) \equiv \frac{-i\; \gamma\cdot p \;A(p^2) + B(p^2)}{p^2 A^2(p^2) + B^2(p^2)} \, ,
\end{equation}
the projected vacuum quark \gls{DSE} in Euclidean space and rainbow truncation can be cast in the form
\begin{subequations}\label{eq:projectedDSE}
\begin{align}
    A(p^2) &= Z_2 + Z_2^2 C_\mathrm{F} \int_k \mathcal{G}(k_\mathrm{G}^2) K_\mathrm{V}(k_\mathrm{G},k_\mathrm{Q}) \sigma_\mathrm{V}(k_\mathrm{Q}^2) \, ,
    \\
    B(p^2) &= Z_4 m(\mu^2) 
            + Z_2^2 C_\mathrm{F} \int_k \mathcal{G}(k_\mathrm{G}^2) K_\mathrm{S}(k_\mathrm{G},k_\mathrm{Q}) \sigma_\mathrm{S}(k_\mathrm{Q}^2) \, ,
\end{align}
\end{subequations}
where $k_\mathrm{G}$ and $k_\mathrm{Q}$ are the gluon and quark momenta in the integral, respectively, $Z_2$ and $Z_4$ are renormalization constants, $C_\mathrm{F}$ is a color factor, and $\mathcal{G}$ an effective interaction.
The reduced propagator projections $K_\mathrm{V,S} = K_\mathrm{V,S}(k_\mathrm{G},k_\mathrm{Q})$ are defined as
\begin{subequations}
\begin{align}
    K_\mathrm{V}(k_\mathrm{G},k_\mathrm{Q}) \sigma_\mathrm{V}(k_\mathrm{Q}^2) 
    &\equiv
        -\frac{i}{4p^2} P_{\mu\nu}^\mathrm{T}(k_\mathrm{G}) \mathrm{Tr}\left[ \gamma \cdot p \gamma_\mu S(k_\mathrm{Q}) \gamma_\nu \right]
         = -\frac{\sigma_\mathrm{V}(k_\mathrm{Q}^2)}{p^2} \left( p\cdot k_\mathrm{Q} + 2 \frac{p\cdot k_\mathrm{G} k_\mathrm{G}\cdot k_\mathrm{Q}}{k_\mathrm{G}^2} \right) \, ,
    \\
    K_\mathrm{S}(k_\mathrm{G},k_\mathrm{Q}) \sigma_\mathrm{S}(k_\mathrm{Q}^2)
    &\equiv
        \frac{1}{4} P_{\mu\nu}^\mathrm{T}(k_\mathrm{G}) \mathrm{Tr}\left[ \gamma_\mu S(k_\mathrm{Q}) \gamma_\nu \right]
         = 3\,\sigma_\mathrm{S}(k_\mathrm{Q}^2) \, ,
\end{align}
\end{subequations}
and the transversal projector is $P_{\mu\nu}^\mathrm{T}(k) = \left( \delta_{\mu\nu} - \frac{k_\mu k_\nu}{k^2} \right)$.
In this parametrization the propagator functions $A$ and $B$ are coupled via the non-linear, in-homogenous integral equations \eqref{eq:projectedDSE}.
With
\begin{equation}
    \Delta_\pm(p) \equiv p A(p^2) \pm i B(p^2)
\end{equation}
Eq.~\eqref{eq:projectedDSE} can be cast in the Form
\begin{equation} \label{eq:DeltapmDSE}
    \Delta_\pm(p) = Z_2 p \pm i Z_4 m + Z_2^2 C_\mathrm{F} \int_k \mathcal{G}(k^2_\mathrm{G})
        \left[ \frac{K_\mp(k_\mathrm{G}, k_\mathrm{Q})}{\Delta_+(k_\mathrm{G})} + \frac{K_\pm(k_\mathrm{G}, k_\mathrm{Q})}{\Delta_-(k_\mathrm{G})} \right] \, ,
\end{equation}
with
\begin{equation}
    K_\pm(k_\mathrm{G}, k_\mathrm{Q})
    \equiv \frac{p}{2k_\mathrm{Q}} K_\mathrm{V}(k_\mathrm{G},k_\mathrm{Q}) \pm \frac{K_\mathrm{S}(k_\mathrm{G},k_\mathrm{Q})}{2}
    =
        -\frac{1}{2pk_\mathrm{Q}} \left( p\cdot k_\mathrm{Q} + 2 \frac{p\cdot k_\mathrm{G} k_\mathrm{G}\cdot k_\mathrm{Q}}{k_\mathrm{G}^2} \right)
        \mp \frac{3}{2} \, ,
\end{equation}
and $p = k_\mathrm{G} + k_\mathrm{Q}$.
Because of $A(p^2), B(p^2) \in \mathds{R}$ for $p^2 \in \mathds{R}^+$ one has $A(p), B(p) \in \mathds{R}$ for $p \in \mathds{R}$ and, due to the Schwarz reflection principle, $A(p) = A^\ast(p^\ast)$ and $B(p) = B^\ast(p^\ast)$; the functions $\Delta_\pm(p)$ are related via
\begin{equation}\label{eq:DeltaRel}
    \Delta_+(p) = \Delta_-^\ast(p^\ast) \, .
\end{equation}
Thus, both functions are related via simple complex conjugation of function and argument.
Therefore, the two coupled Eqs.~\eqref{eq:projectedDSE} are equivalent to
\begin{equation} \label{eq:DeltaDSE}
    \Delta(p) = Z_2 p + i Z_4 m + Z_2^2 C_\mathrm{F} \int_k \mathcal{G}(k^2_\mathrm{G})
        \left[ \frac{K_-(k_\mathrm{G}, k_\mathrm{Q})}{\Delta(k_\mathrm{G})} + \frac{K_+(k_\mathrm{G}, k_\mathrm{Q})}{\Delta^\ast(k^\ast_\mathrm{G})} \right] \, ,
\end{equation}
where our choice is $\Delta(p) \equiv \Delta_+(p)$.
Instead of dealing with two functions coupled via the integral equations \eqref{eq:projectedDSE}, it suffices to solve Eq.~\eqref{eq:DeltaDSE}.
The key is that the two non-equivalent functions in question, $A$ and $B$, are only coupled via the two equations \eqref{eq:projectedDSE}.
In contrast, the two functions coupled by Eq.~\eqref{eq:DeltapmDSE} are also related by complex conjugation of function and argument.
All \emph{``traditional''} propagator functions can uniquely be determined from $\Delta(p)$,
\begin{subequations}
\begin{align}
    A(p^2) &= \frac{1}{2p} \left( \Delta(p) + \Delta^\ast(p^\ast) \right) \, ,
    &
    B(p^2) &= \frac{1}{2i} \left( \Delta(p) - \Delta^\ast(p^\ast) \right) \, ,
    \\
    \sigma_\mathrm{V}(p^2) &= \frac{1}{2p} \left( \frac{1}{\Delta(p)} + \frac{1}{\Delta^\ast(p^\ast)} \right) \, ,
    &
    \sigma_\mathrm{S}(p^2) &= \frac{i}{2} \left( \frac{1}{\Delta(p)} - \frac{1}{\Delta^\ast(p^\ast)} \right) \, .
\end{align}
\end{subequations}
For $p \in \mathds{R}$ the two real functions $A$ and $B$ are replaced by one complex function $\Delta$.
However, for $p \in \mathds{C}$, $A$ and $B$ are complex and, according to Eq.~\eqref{eq:DeltaDSE}, equivalent to the information contained in one complex function.
Due to the symmetries of the propagator functions w.\,r.\,t.\ complex conjugation this is not completely surprising.

Eq.~\eqref{eq:DeltaDSE} may also be used to study the analytic properties of the quark propagator in the complex plane.
If $\Delta_+(p_+) = 0$ or $\Delta_-(p_-) = 0$ and, hence,
\begin{equation}\label{eq:DeltaRoot}
    p_\pm = \mp i \frac{Z_4}{Z_2} m - Z_2 C_\mathrm{F} \int_k \mathcal{G}(k^2_\mathrm{G})
        \left[ \frac{K_\mp(k_\mathrm{G}, k_\mathrm{Q})}{\Delta_\pm (k_\mathrm{G})} + \frac{K_\pm (k_\mathrm{G}, k_\mathrm{Q})}{\Delta_\mp(k_\mathrm{G})} \right] \, ,
\end{equation}
the dressing functions $\sigma_\mathrm{V}$ and $\sigma_\mathrm{S}$ have poles.
Note, that $p$ enters the integral on the r.\,h.\,s., i.\,e., this equation requires a self-consistent solution.
Because of Eq.~\eqref{eq:DeltaRel}, one has $p_+ = p_-^\ast$, and, hence, Eq.~\eqref{eq:DeltaRoot} can further be utilized to give two separate equations for the poles of the quark propagator in the complex plane.

\section{Essentials of the Bethe-Salpeter equation}
\label{sct:BSE}
The \gls{DSBSE} approach to hadrons uses the \gls{BSE} to describe mesons as quark-antiquark systems \cite{Bethe:1951bs}, a covariant Faddeev equation for a three-quark setup to study baryons \cite{Eichmann:2009qa,Nicmorus:2008vb,Nicmorus:2008eh}, and more complicated equations to go beyond these configurations \cite{Bicudo:2001jq,Cotanch:2002vj,Biernat:2014xaa,Eichmann:2015cra}. The covariant amplitudes are then used for the subsequent calculation of form factors or other transiton amplitudes \cite{Maris:1999bh,Bloch:2003vn,Holl:2005vu,Eichmann:2007nn,Eichmann:2008ef,Mader:2011zf,Sanchis-Alepuz:2013iia,Sanchis-Alepuz:2014wea,Biernat:2015xya,Bedolla:2015mpa,Bedolla:2016yxq,Raya:2016yuj}. For our discussion herein, we only sketch the essential aspects related to the homogenous quark-bilinear \gls{BSE}.

In the quark model, bound states are constructed from their constituents to represent a particular state with well-defined quark orbital-angular momentum $l$ and total quark spin $s$.
The possible set of states with total angular momentum $J$, parity $\mathcal{P}$ and, if the state is its own anti-particle via charge conjugation, charge conjugation parity $\mathcal{C}$ is then limited by
\begin{equation}
    |l-s|\le J \le |l+s| \,,\quad \mathcal{P}=(-1)^{l+1}\,,\quad \mathcal{C}=(-1)^{l+s}\, .
\end{equation}
In a way the quantum numbers $J^{\mathcal{P(C)}}$ are determined \emph{a posteriori} and constrained, in particular, via adjusting the quark orbital-angular momentum $l$.
However, the latter is neither observable, nor a Poincar\'e covariant quantity, and, thus, actually not a good quantity to characterize a state initially.

Within a \gls{BSE} approach the construction of a bound state follows the opposite path.
A particular state is characterized by the Poincar\'e covariant quantum numbers $J^{\mathcal{P(C)}}$ ($\mathcal{C}$ of course only when it is an eigenstate of the charge conjugation operator), which determines the basis elements to construct the bound state's \gls{BSA} \cite{Krassnigg:2010mh}.
At this point, any particular Lorentz frame may be chosen. While choosing a non-restframe can be advantageous numerically in the calculation of transitions that actually involve boosted amplitudes \cite{Bhagwat:2006pu}, any standard meson-spectrum calculation is sensibly be performed in the meson's rest frame.
The quark orbital-angular momentum $l$ may then be determined by projecting the bound state's \gls{BSA} onto eigenstates of the Pauli-Lubanski operator and evaluating their contributions to the canonical norm of the \gls{BSA} \cite{Bhagwat:2006xi,Hilger:2015ora}.
In this way, one naturally includes all possible wave contributions including ones that are impossible in the quark model, and, in particular, also automatically generates mixed wave contributions, as the norm is bilinear in the wave function.
Thus, the \gls{DSBSE} approach to meson bound states provides a natural means to investigate and discuss the partial-wave characteristics of a state by virtue of its manifest and inherent Poincar\'e covariance; notably, also those characteristics may be obtained, which have no correspondence in non-Poincar\'e covariant physics.

In terms of the computational setup, a crucial strategy in order to keep the full Poincar\'e covariance of the approach is to perform the full hyperpolar angle integration instead of expanding the \gls{BSA}'s angular dependence in, e.\,g., Gegenbauer \cite{UweHilger:2012uua} or Chebyshev \cite{Krassnigg:2003dr} polynomials.
As a result the dependence of the bound-state mass on the momentum partitioning parameter, which can be considerable in case of an expansion \cite{Alkofer:2002bp}, is below the numerical precision, i.\,e., when the bound state mass is determined up to a particular tolerance, the variation w.\,r.\,t.\ the momentum partitioning parameter remains below this tolerance.
Therefore, one may fully exploit the choice of the momentum partitioning parameter in order to keep non-analyticities of the quark propagator outside of the \gls{BSE}'s integration domain as far as possible \cite{UweHilger:2012uua,Hilger:2016efh}.

\section{Results and discussion}
\label{sct:results}

\begin{figure}[t]
\centering
\includegraphics[width=0.48\textwidth,clip]{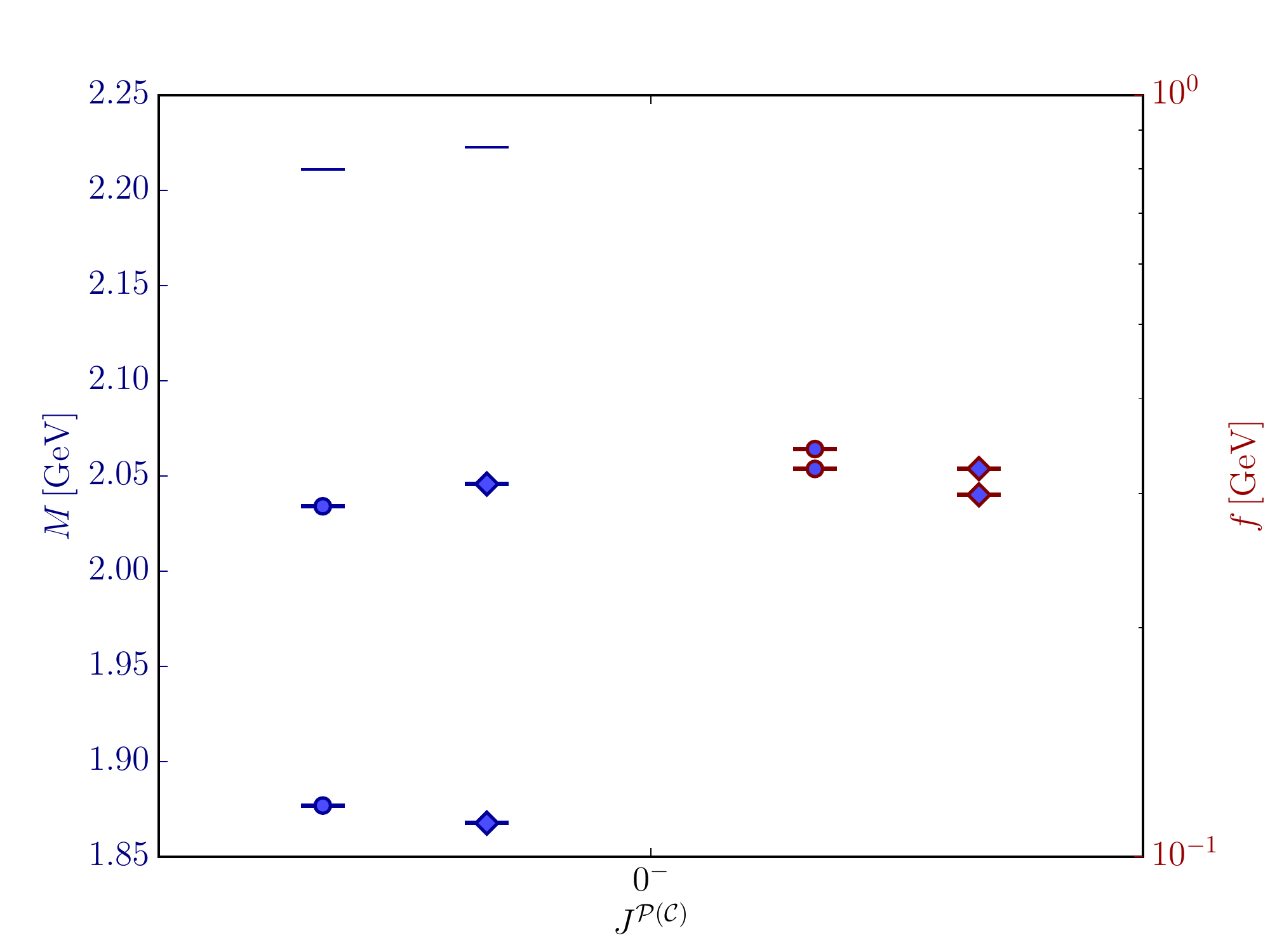}
\includegraphics[width=0.48\textwidth,clip]{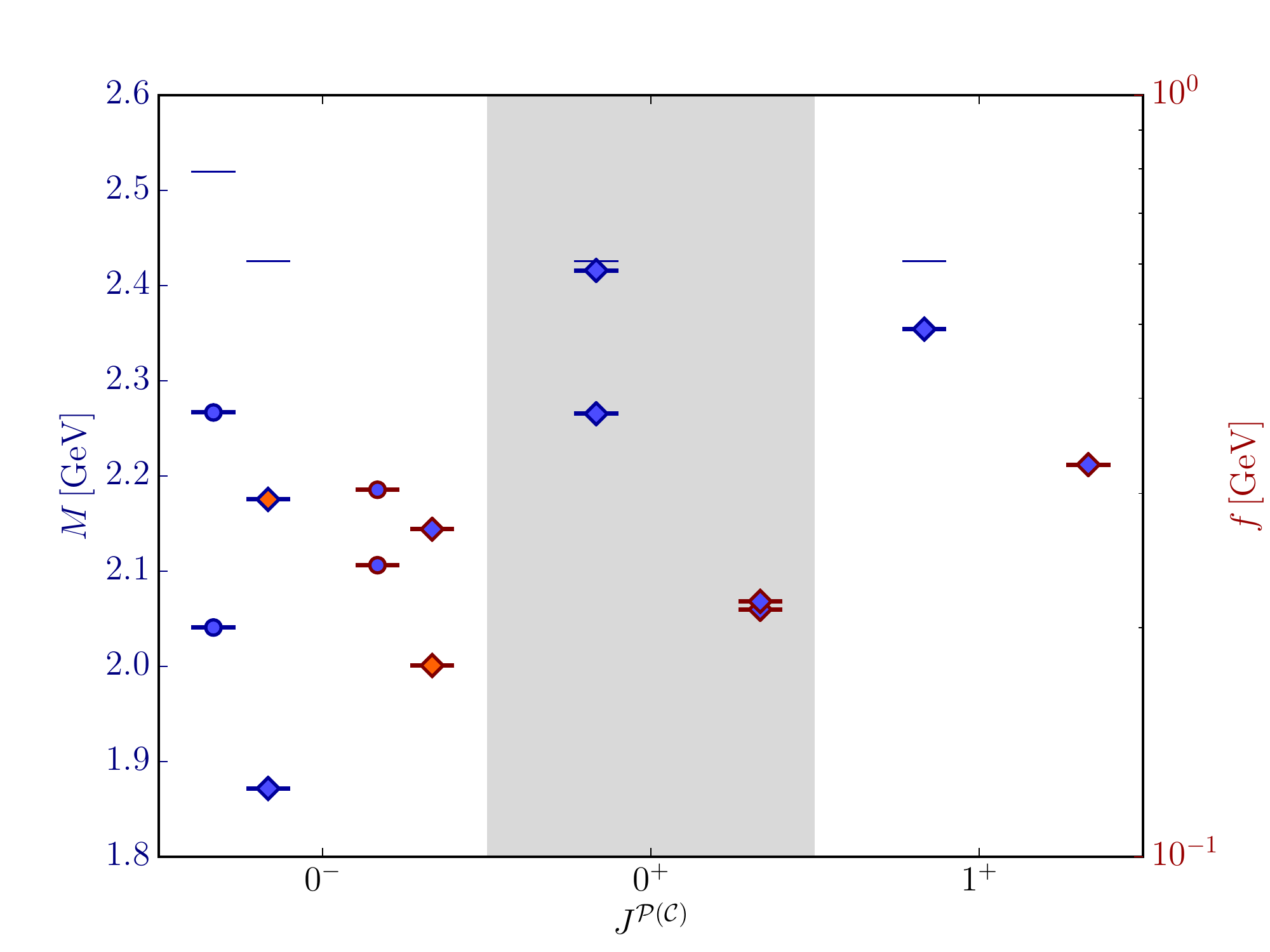}
\includegraphics[width=0.48\textwidth,clip]{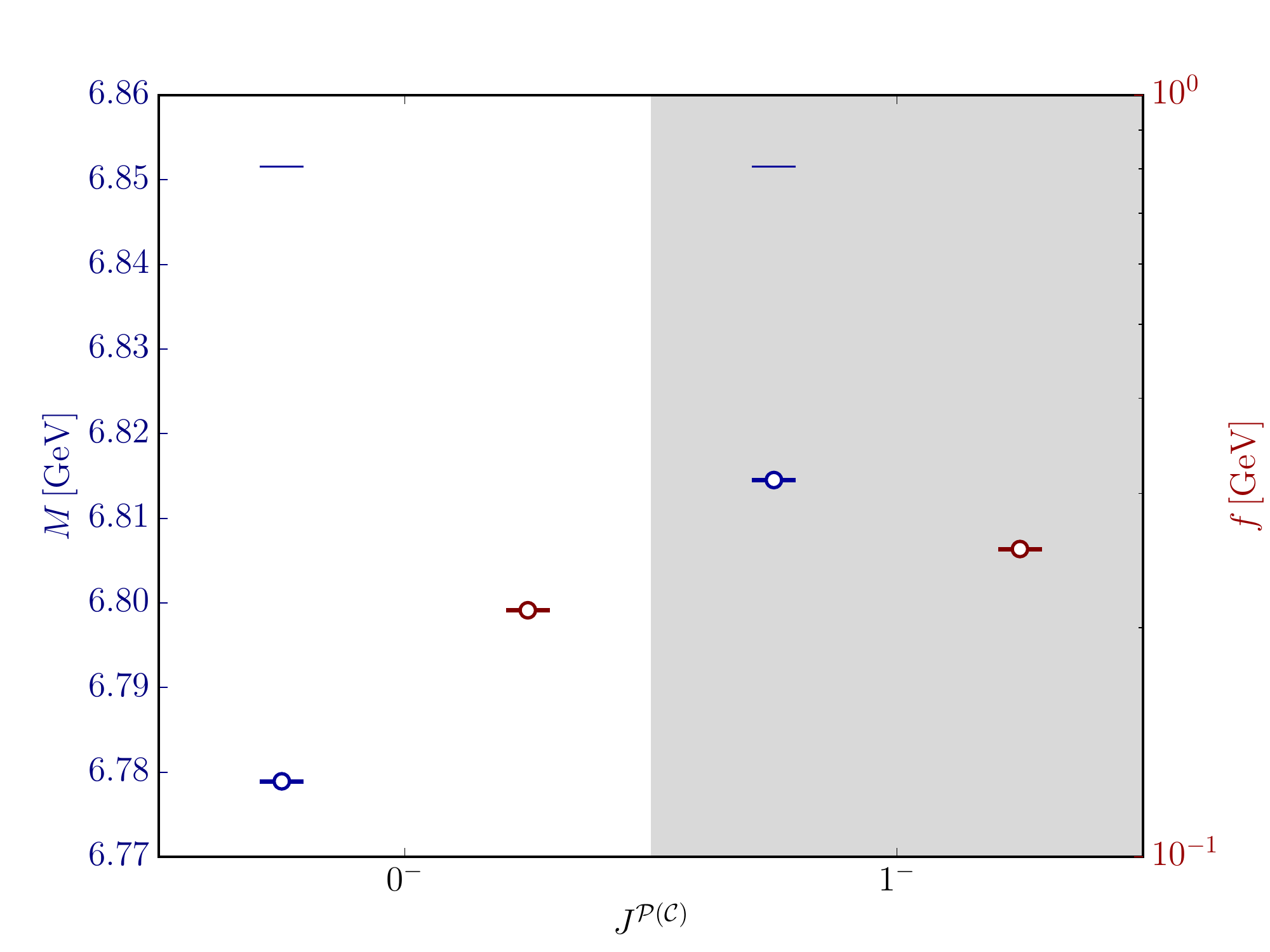}
\includegraphics[width=0.48\textwidth,clip]{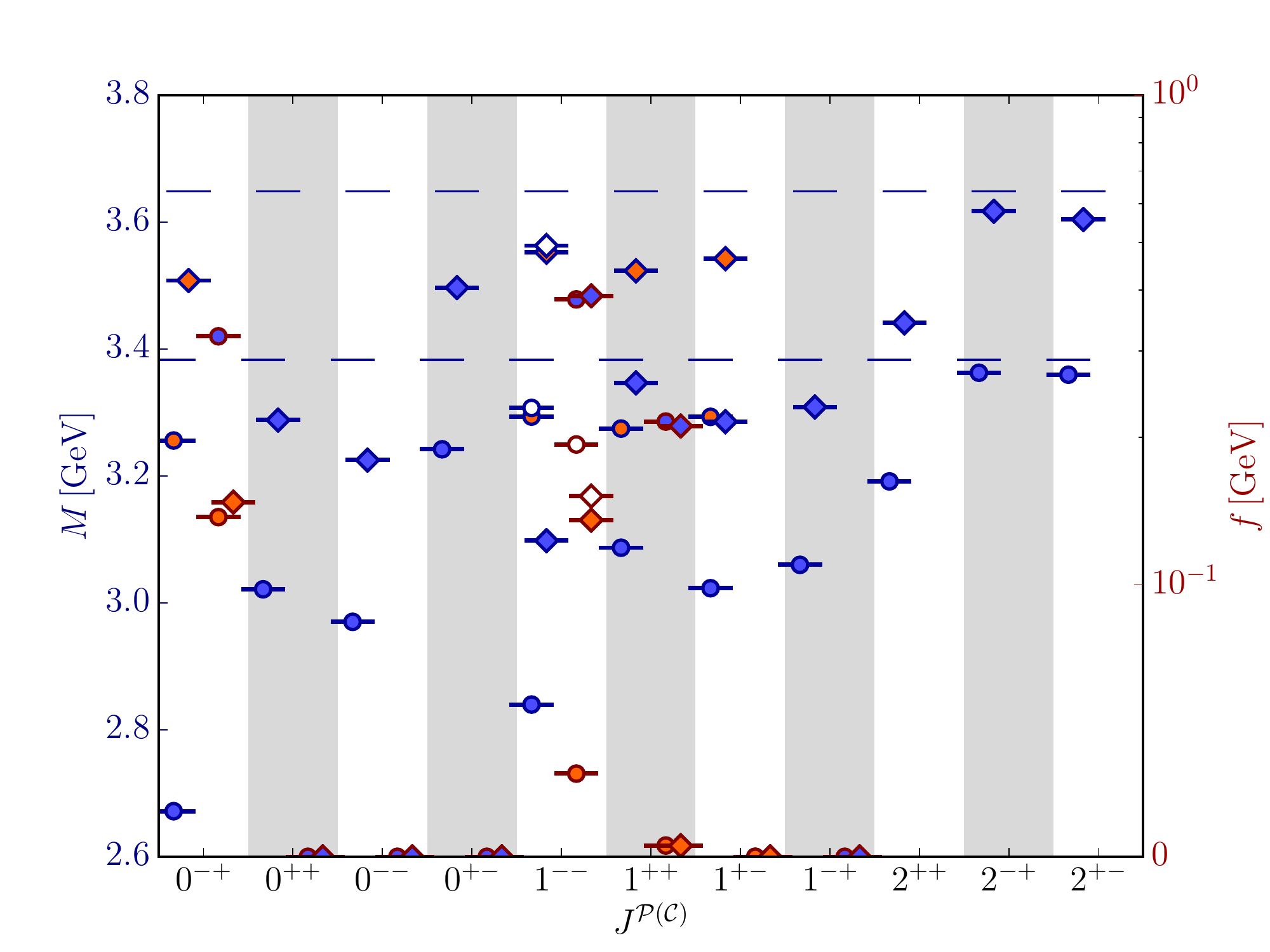}
\caption{
Results for $\bar q c$ (upper left panel), $\bar s c$ (upper right panel), $\bar c b$ (lower left panel) and
$\bar c c$ (lower right panel). Meson bound state spectrum (blue and left axis) and leptonic decay constants (red and right axis) for all quantum numbers $J^{\mathcal{P(C)}}$ where bound states have been found below the pole threshold (short horizontal lines). For description of marker shape, fill and border colors see text. Red scale is linear below $10^{-1}\,\mathrm{GeV}$.
}
\label{fig:qc&sc}
\end{figure}

In order to investigate meson properties herein we use a well-known effective interaction \cite{Alkofer:2002bp} with the model parameters $\omega=0.5$ GeV and $D=1/\omega^4$ GeV${}^{-2}$. The results for the meson masses as well as leptonic decay constants for the following quark-flavor combinations are presented in Fig.~\ref{fig:qc&sc}: $\bar q c$ (upper left panel), $\bar s c$ (upper right panel), $\bar c b$ (lower left panel), and $\bar c c$ (lower right panel). To avoid one particular fitting bias we adjust the current quark masses for the charm quark in two different ways: on the one hand, we aim at the experimental $D$-meson mass, and on the other hand, we aim at the experimental mass of the $J/\Psi$, both well known quantities. 

The former set of results is presented by filled circles in all panels of Fig~\ref{fig:qc&sc}, the latter set is represented by filled diamonds. The sets are presented next to each other for easy comparison. In each plot, the left (blue) axis corresponds to the meson mass $M$, while the right (red) axis corresponds to its leptonic decay constant $f$, which is realized in the data points by their boundary color. In addition, excitations are coded via the fill color of the symbol such that blue filling represents ground states, red corresponds to the first excitation, and white to a second excitation. Furthermore, for each set and plot, short horizontal lines are plotted to indicate the pole threshold discussed above.

\begin{figure}[t]
\centering
\includegraphics[width=7cm,clip]{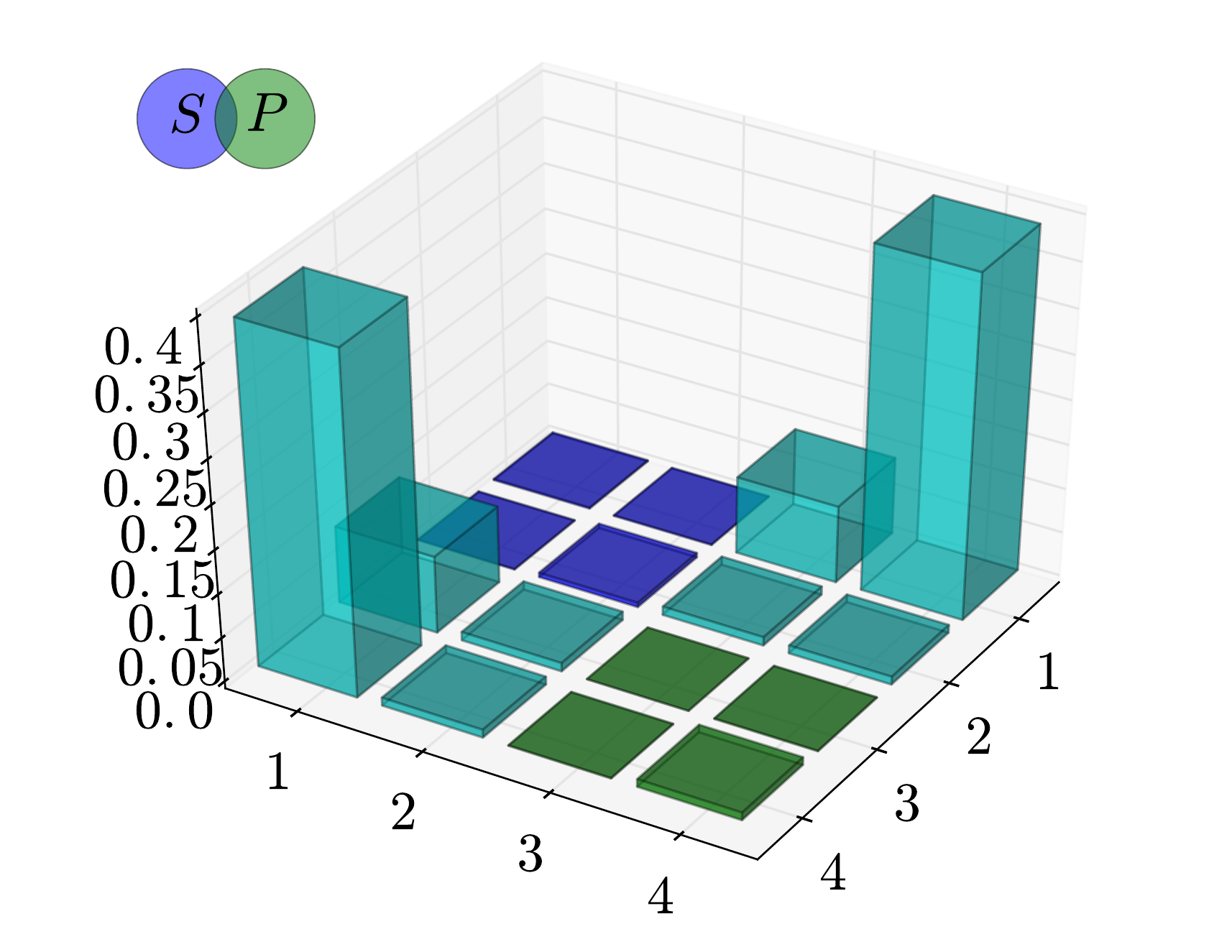}
\includegraphics[width=7cm,clip]{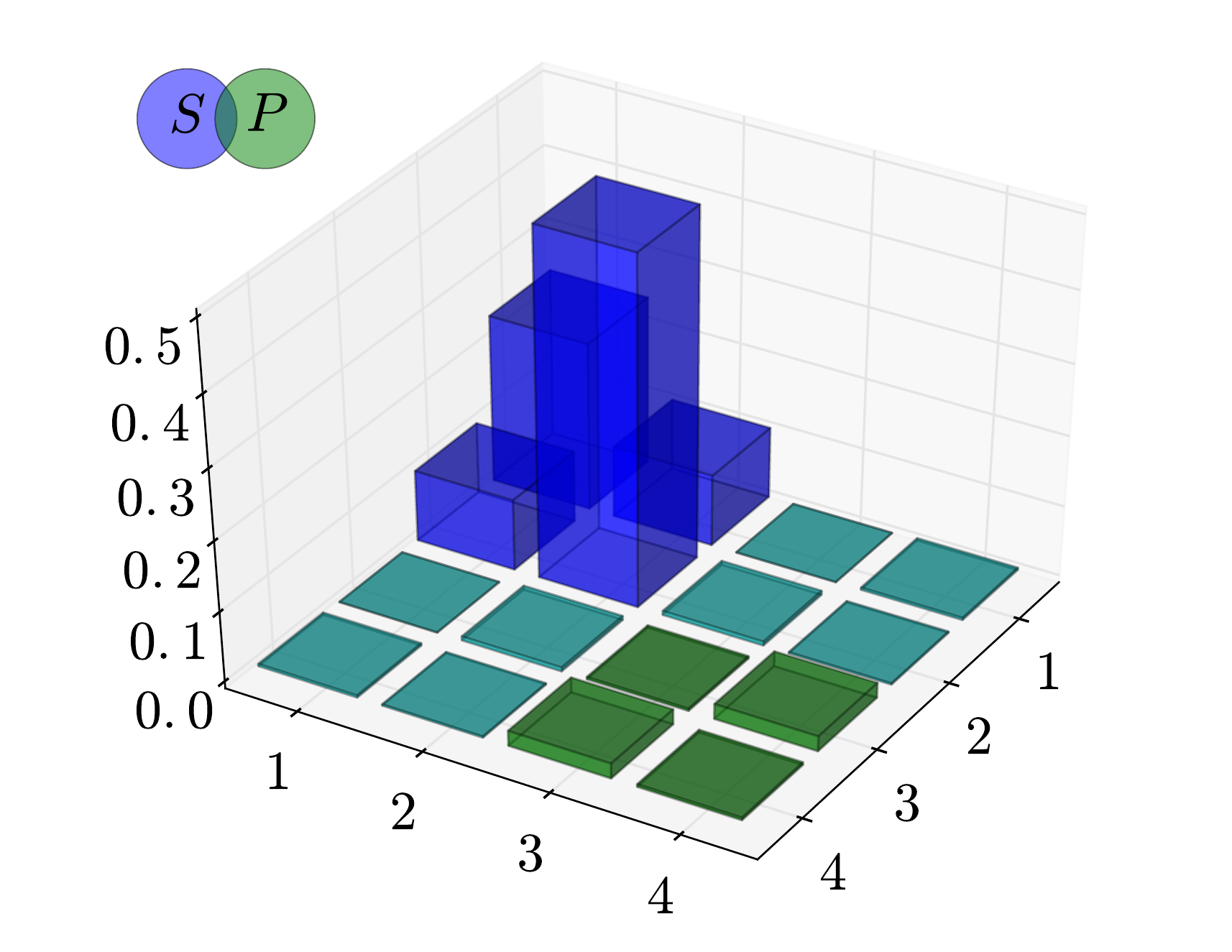}
\includegraphics[width=7cm,clip]{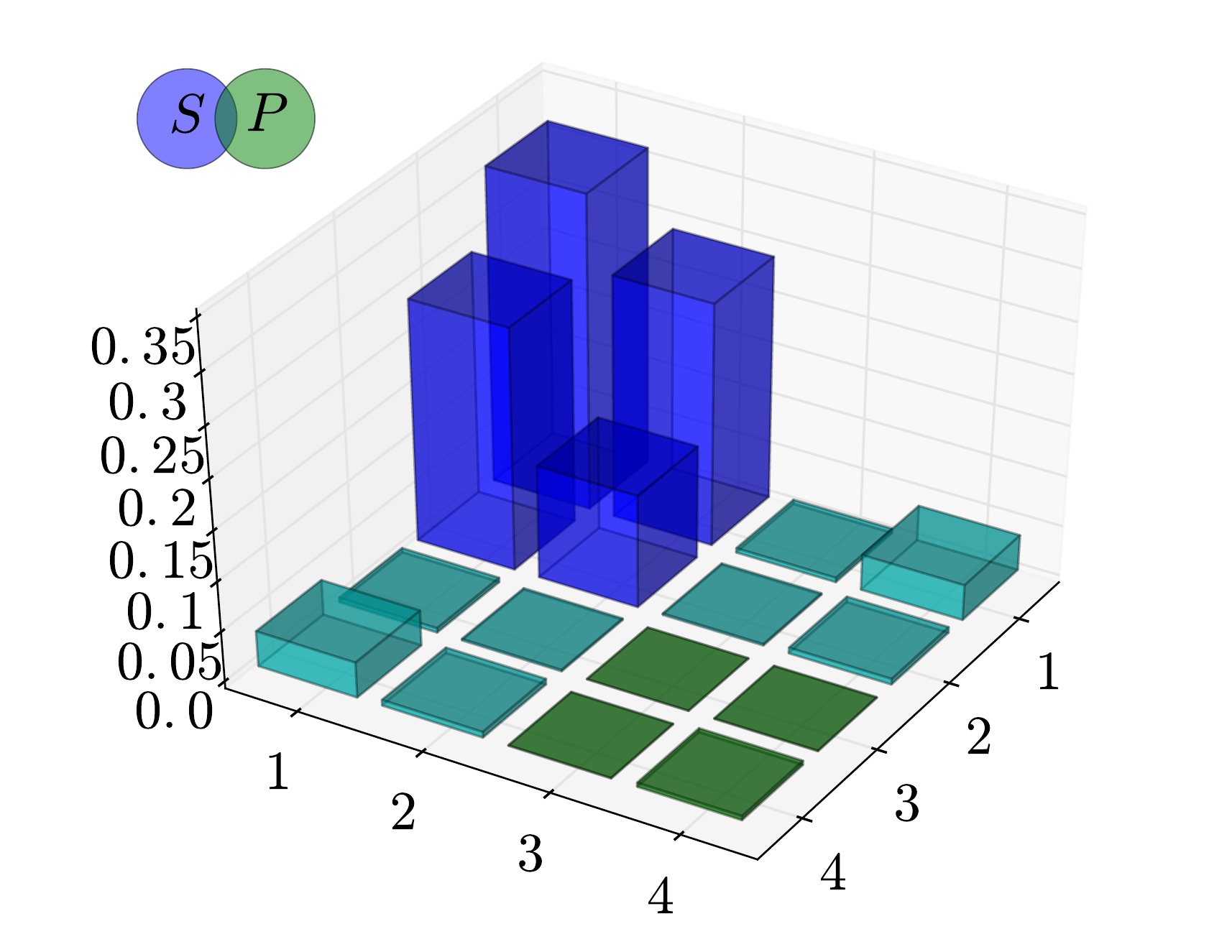}
\includegraphics[width=7cm,clip]{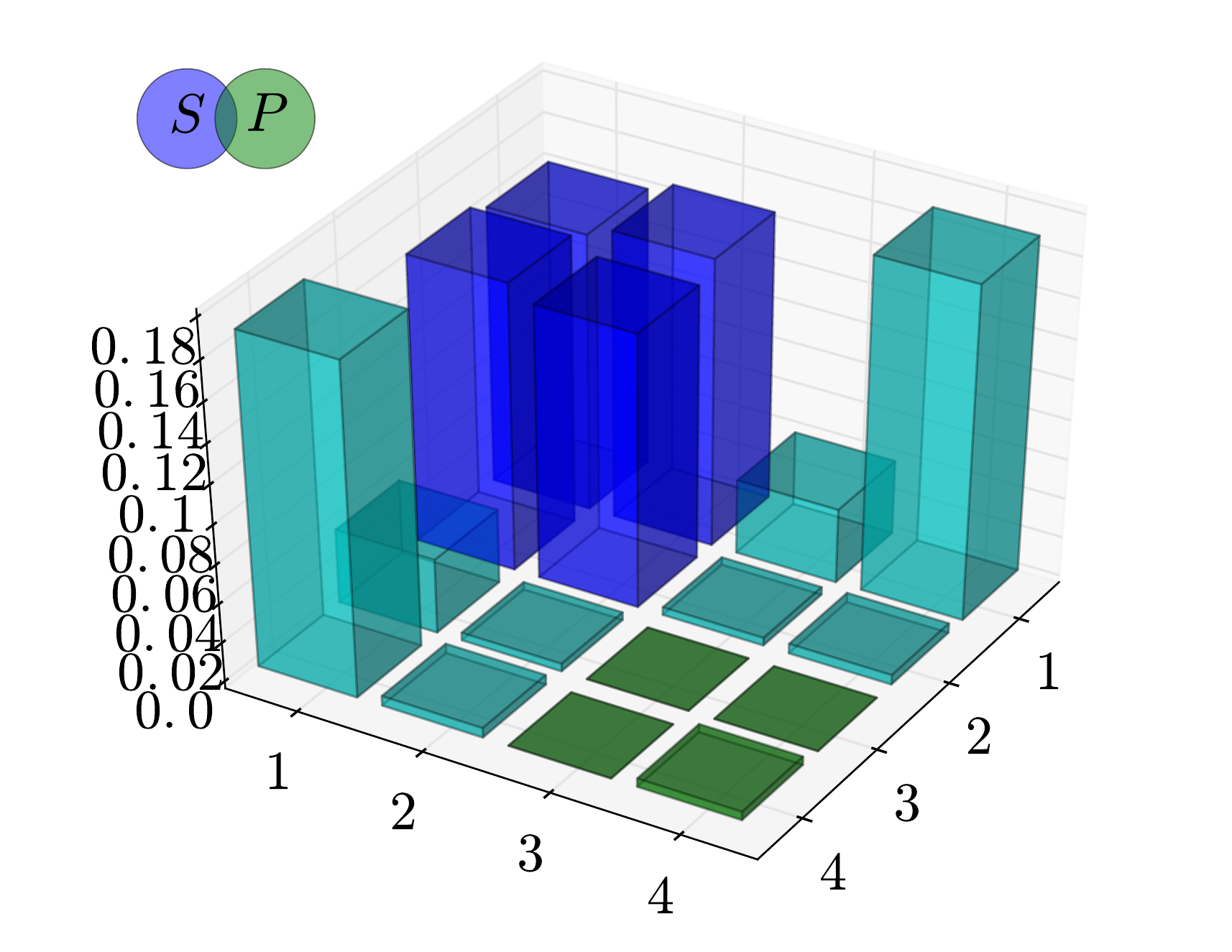}
\caption{
\gls{OAMD} of the pseudoscalar $\mathrm{D_s}$ mesons: ground (upper left panel), first excited (upper right panel), second excited (lower left panel), and alternative ground state; plotted are the squared contributions of all covariant combinations to the canonical norm of the \gls{BSA}, normalized to $1$.
Numbers along $x-$ and $y-$ axes identify covariants in \gls{BSA} basis, see, e.\,g., \cite{Hilger:2015ora}.
}
\label{fig:OAMD_Ds}
\end{figure}

Overall we find a fairly rich spectrum with a reasonable number of excitations below the pole threshold, i.\,e., which can thoroughly be evaluated on-shell without extraction from off-shell data and with reliable numerical stability.
However, apart from the pole threshold, one immediately encounters two other issues which impede the investigation and reduce the number of accessible states when approaching the open-flavor meson spectrum within the \gls{DSBSE} approach.

First of all, one may find non-monotonic behavior of the eigenvalue curves w.\,r.\,t.\ the bound state momentum $P^2$, which are monotonically growing functions for quarkonium states.
In such a case, the existence of a solution may depend strongly on details of the model or parameter values, since any small influence can shift a curve enough to avoid intersecting $1$, which is the condition for a solution of the homogeneous equation when it is solved as an eigenvalue problem \cite{Krassnigg:2003wy}.
One then faces the simple numerical problem to find a quark mass for which solutions exist, which often requires manual intervention or a very fine bound state momentum $P^2$ grid.
If the latter is too coarse, the intersection with unity might be missed for solution strategies relying on eigenvalues as well as the corresponding determinant.
In addition, neighboring solutions along the same eigenvalue curve can occur like, e.\,g., for the $\bar q c$ states in Fig.~\ref{fig:qc&sc} (upper left panel).

Secondly, two neighbouring eigenvalue curves can, upon crossing, collapse on a certain domain in the sense that they have identical real parts, but obtain non-zero complex conjugated imaginary parts.
Clearly, there is no way to obtain a bound state solution in a case, where the real part of such a pair of complex-valued curves intersects $1$.
One might, however, suspect that solutions exist for complex bound-state momentum squared.
On the other hand, since the \gls{RL} truncated \gls{BSE} does not include hadronic decay channels, which one would expect when finding a decay width of the state, it is not clear how to interpret such a scenario. Note that an iterative procedure in finding the eigenvalues \cite{Krassnigg:2003wy} cannot produce complex conjugate eigenvalue pairs correctly; more sophisticated methods are needed to obtain correct results in this case \cite{Blank:2011qk,Blank:2010bp}.

Recently, we have performed an analysis of the \gls{OAMD} of \gls{BSA}s for quarkonia as well as open-flavor states with strangeness \cite{Hilger:2015ora,Hilger:2016efh}. 
Since a charmed or any other heavy-light meson is more unbalanced than a kaon, the question arises how the \gls{OAMD} looks for the states considered here.
As an example, we depict the result of our \gls{OAMD} analysis of the $\mathrm{D_s}$ ground- and excited states in Fig.~\ref{fig:OAMD_Ds} as a continuation of the studies in \cite{Hilger:2015ora} and \cite{Hilger:2016efh} with the corresponding effective interaction of \cite{Maris:1999nt}.

A standard equal-flavor pseudoscalar meson is predominantly an $\mathrm{S}$-wave (blue) state with only minor or even negligible mixed $\mathrm{S-P}$-wave and $\mathrm{P}$-wave contributions, well in accord with expectations from the quark model \cite{Hilger:2015ora}.
With different-mass quark ingredients, however, this can change. 
For the depicted pseudoscalar $\mathrm{D_s}$ ground state we find a predominantly mixed configuration with negligible pure $\mathrm{S}$- and $\mathrm{P}$-wave contributions, which is rather unusual. To investigate this, we redid the analysis for the ground state with a slightly shifted strange-quark mass; the result, which is more along the typical lines, is shown in the lower right panel of Fig.~\ref{fig:OAMD_Ds}. Apparently, the unusual ground-state result is not robust, and further investigation is needed to clarify the situation of \gls{OAMD} in heavy-light states.
Interestingly, both excitations, the first of which is a quasi-exotic state, follow the pattern of being predominantly $\mathrm{S}$-wave states.


\section{Summary and outlook}
\label{sct:summary}

In summary, we have demonstrated that the coupled vacuum equations for the quark propagator dressing functions can be cast to a single equation for a single function which can uniquely be mapped to the original propagator functions.
While for $p^2 \geq 0$ the numerical effort stays the same, it is considerably reduced for $p^2 \in \mathds{C}\setminus\mathds{R}^+$.

Furthermore, we have presented heavy-light meson phenomenology and also revealed the major challenges and technical issues to be addressed in further open-flavor studies within the combined \gls{DSBSE} approach. Several aspects, as described above, can either obstruct a direct investigation of the higher-lying spectrum or distort the expected behavior of one or a pair of eigenvalue curves of the \gls{BSE} kernel; all of these deserve further study.

Finally, we performed the \gls{OAMD} for ground state and two excitations in the $\mathrm{D_s}$ channel.
While we get an unusual result for the ground state, closer analysis shows that also charmed strange pseudoscalars appear to be mostly $\mathrm{S}$-wave states.

Next, the investigation will be completed by including all possible quark flavor combinations and all accessible quantum numbers $J^{\mathcal{P(C)}}$.
This will make our study the most comprehensive investigation of the meson spectrum and leptonic decay constants in terms of quark flavors and quantum numbers $J^{\mathcal{P(C)}}$ currently available in the \gls{DSBSE} approach, providing a benchmark until the above mentioned issues have been resolved.

\section*{Acknowledgement}
This work was supported by the Austrian Science Fund (FWF) under project no.\ P25121-N27.
We acknowledge discussions with M.\,Gomez-Rocha, G.\,Eichmann, R.\,Williams, B.\,K\"ampfer, S.\,Dorkin, L.\,Kaptari and R.\,Greifenhagen.


\input{ProcCONF16.bbl}

\end{document}

%% file: glossaries.tex
\newacronym{OAMD}{OAMD}{orbital-angular momentum decomposition}
\newacronym{DS}{DS}{Dyson-Schwinger}
\newacronym{BS}{BS}{Bethe-Salpeter}
\newacronym{MT}{MT}{Maris-Tandy}
\newacronym{AWW}{AWW}{Alkofer-Watson-Weigel}
\newacronym{NG}{NG}{Nambu-Goldstone}
\newacronym{WW}{WW}{Wigner-Weyl}
\newacronym{SM}{SM}{Standard Model}
\newacronym{DSE}{DSE}{Dyson-Schwinger equation}
\newacronym{BSE}{BSE}{Bethe-Salpeter equation}
\newacronym{DSBS}{DSBS}{Dyson-Schwinger--Bethe-Salpeter}
\newacronym{DSBSE}{DSBSE}{Dyson-Schwinger-Bethe-Salpeter-equation}
\newacronym{BSA}{BSA}{Bethe-Salpeter amplitude}
\newacronym{BSM}{BSM}{Bethe-Salpeter matrix}
\newacronym{BSV}{BSV}{Bethe-Salpeter vertex}
\newacronym{QED}{QED}{quantum electrodynamics}
\newacronym{QCD}{QCD}{quantum chromodynamics}
\newacronym{QCDg}{QCD}{Quantenchromodynamik}
\newacronym{DCSB}{D$\upchi$SB}{dynamical chiral symmetry breaking}
\newacronym{VMD}{VMD}{vector meson dominance}
\newacronym{FAIR}{FAIR}{Facility for Antiproton and Ion Research}
\newacronym{HIC_for_FAIR}{HIC for FAIR}{Helmholtz International Center for FAIR}
\newacronym{WASAatCOSY}{WASA at COSY}{Wide Angle Shower Apparatus at Cooler Synchrotron}
\newacronym{WASA}{WASA}{Wide Angle Shower Apparatus}
\newacronym{COSY}{COSY}{Cooler Synchrotron}
\newacronym{MAMI}{MAMI}{\emph{Mainzer Mikrotron}}
\newacronym{LHC}{LHC}{Large Hadron Collider}
\newacronym{LHCb}{LHC{\scshape b}}{Large Hadron Collider beauty experiment}
\newacronym{KLOE}{KLOE}{K Long Experiment}
\newacronym{DAFNE}{DA$\Phi$NE}{Double Annular $\Phi$ Factory for Nice Experiments}
\newacronym{DESY}{DESY}{\emph{Deutsches Elektronen-Synchrotron}}
\newacronym{JLab}{JLab}{Thomas Jefferson National Accelerator Facility}
\newacronym{QSR}{QSR}{QCD sum rule}
\newacronym{OPE}{OPE}{operator product expansion}
\newacronym{JINR}{JINR}{Joint Institute for Nuclear Research}
\newacronym{HZDR}{HZDR}{Helmholtz-Zentrum Dresden-Rossendorf}
\newacronym{ANL}{ANL}{Argonne National Laboratory}
\newacronym{PANDA}{PANDA}{facility for anti-Proton ANnihilation at DArmstadt}
\newacronym{EFT}{EFT}{effective field theory}
\newacronym{VEV}{VEV}{vacuum expectation value}
\newacronym{KFU}{KFU}{Karl-Franzens University}
\newacronym{lQCD}{{\scshape l}QCD}{lattice-regularized QCD}
\newacronym{WTI}{WTI}{Ward-Takahashi identity}
\newacronym{AVWTI}{{\scshape av}WTI}{axialvector Ward-Takahashi identity}
\newacronym{RL}{RL}{rainbow ladder}
\newacronym{TUD}{TUD}{Technical University of Dresden}
\newacronym{UU}{UU}{Uppsala University}
\newacronym{NJL}{NJL}{Nambu--Jona-Lasinio model}
\newacronym{QFT}{QFT}{quantum field theory}
\newacronym{BRST}{BRST}{Becchi-Rouet-Stora-Tyutin}
\newacronym{FAKT}{FAKT}{\emph{``Fachausschuss Kern- und Teilchenphysik''}}
\newacronym{OPG}{{\"O}PG}{\emph{``{\"O}sterreichische Physikalische Gesellschaft''}}
\newacronym{APS}{{\"O}PG}{Austrian Physical Society}
\newacronym{SLAC}{SLAC}{Stanford Linear Accelerator Center}
\newacronym{DPG}{DPG}{\emph{``Deutsche Physikalische Gesellschaft''}}
\newacronym{GPS}{DPG}{German Physical Society}
\newacronym{FWF}{FWF}{\emph{``Fonds zur F{\"o}rderung der wissenschaftlichen Forschung''}}
\newacronym{KSU}{KSU}{Kent State University}
\newacronym{GMOR}{GMOR}{Gell-Mann--Oakes--Renner}
\newacronym{QGV}{QGV}{quark-gluon interaction vertex}
\newacronym{STI}{STI}{Slavnov-Taylor identity}
\newacronym{GU}{GU}{Giessen University}
\newacronym{GSI}{GSI}{\emph{Gesellschaft f{\"u}r Schwerionenforschung}}
\newacronym{ERA}{ERA}{European Research Area}
\newacronym{BRL}{BRL}{beyond-rainbow-ladder}
\newacronym{WP}{WP}{work package}
\newacronym{D}{D}{major deliverable}
\newacronym{BEPC}{BEPC}{Beijing Electron Positron Collider}
\newacronym{BES}{BES-III}{Bejing Spectrometer}
\newacronym{KEK}{KEK}{High Energy Accelerator Research Organization}